\documentclass[%
 reprint,
 amsmath,amssymb,
 aps,
]{revtex4-2}
\usepackage{graphicx}
\usepackage{xcolor}   % color の代わりに xcolor でもOK
\usepackage{dcolumn}   % needed for some tables
\usepackage{bm}        % for math
\usepackage{amssymb}   % for math
\usepackage{setspace} 
\usepackage{here}
\usepackage{float}

\begin{document}

\newcommand{\ddt}[1]{{\frac{\partial {#1}}{\partial t}}}
\newcommand{\dzt}[1]{{\frac{d {#1}}{d t}}}
\newcommand{\bi}[1]{{{\textbf{\emph{#1}}}}}
\newcommand{\bd}[1]{{{\textbf{#1}}}}
\newcommand{\el}[0]{{{_\textrm{e}}}}
\newcommand{\ddx}[1]{{\frac{\partial {#1}}{\partial x}}}
\newcommand{\ddz}[1]{{\frac{\partial {#1}}{\partial z}}}
%----------------------------------------------------------------------------------------

\preprint{APS/123-QED}
\title{Relativistic multi-stage resonant and trailing-field acceleration induced by large-amplitude Alfv\'en waves in a strong magnetic field}
\author{S. Isayama} \affiliation{Faculty of Engineering Sciences, Kyushu University, 6-1 Kasuga-Koen, Kasuga, Fukuoka 816-8580, Japan}
\affiliation{International Research Center for Space and Planetary Environmental Science (i-SPES), Kyushu University, Motooka, Nishi-Ku, Fukuoka 819-0395, Japan}
\author{T. Sano} \affiliation{Institute of Laser Engineering, Osaka University, Suita, Osaka 565-0871, Japan} 
\author{S. Matsukiyo} \affiliation{Faculty of Engineering Sciences, Kyushu University, 6-1 Kasuga-Koen, Kasuga, Fukuoka 816-8580, Japan}
\affiliation{International Research Center for Space and Planetary Environmental Science (i-SPES), Kyushu University, Motooka, Nishi-Ku, Fukuoka 819-0395, Japan}
\author{S. H. Chen} \affiliation{Department of Physics, National Central University, Jhongli District, Taoyuan 32001, Taiwan}

\date{\today}% It is always \today, today,
             %  but any date may be explicitly specified

\begin{abstract}
We propose a novel particle acceleration mechanism driven by large-amplitude Alfv\'en waves in a strong magnetic field. The acceleration process proceeds through multiple stages triggered by counter propagating wave-particle resonant acceleration (CWRA) via decay instability. Initially, parent and daughter Alfv\'en waves resonantly accelerate particles perpendicular to the ambient magnetic field. The resultant modulational instability generates electrostatic fields within the wave packet, which are locally amplified by the ponderomotive force of the Alfvén wave packet. These fields subsequently drive further acceleration within a few relativistic gyroperiods via gyroresonant surfing acceleration (GRSA). During this, the $\mathbf{v} \times \mathbf{B}$ force facilitates momentum transfer from the perpendicular to the parallel direction. In the later stage, particles become trapped by the parent wave and gain additional energy through single wave resonant acceleration (SWRA). Furthermore, the accumulation of accelerated particles induces electrostatic trailing fields behind and at the tail of the wave packet, which drive trailing-field acceleration (TFA) of other electrons. The combined effects of these mechanisms, CWRA followed by GRSA and SWRA, result in highly relativistic electron energy. The electron energy accelerated through the above process is higher than that accelerated through TFA. This multi-stage acceleration process provides new insights into the generation of high energy cosmic rays in astrophysical environments.

%Magnetowave-induced plasma wakefield acceleration (MPWA) in a relativistic astrophysical outflow has been proposed as a possible mechanism for the acceleration of high-energy cosmic rays. However, relativistic particle acceleration by this acceleration mechanism has not yet been demonstrated. Here, we propose the wakefield excitation mechanism in a strongly magnetized via cycrotlon resonance. that wakefield acceleration becomes possible via cyclotron resonance when the wave amplitude of the Alfv\'en wave packet exceeds a critical value. Furthermore, we show that relativistic resonant particle acceleration is the dominant acceleration process in a propagating Alfv\'en wave packets.
\end{abstract}

\pacs{Valid PACS appear here}% PACS, the Physics and Astronomys
                             % Classification Scheme.
\maketitle

%\tableofcontents

\section{Introduction}
The acceleration mechanism of high energy cosmic rays (CRs) has been a long standing mystery in astrophysics. We focus in this study on the particle acceleration through the interactions between charged particles and coherent Alfv\'{e}n/whistler waves under strong ambient magnetic field. The quasi-periodic oscillations of the emissions from soft gamma-ray repeaters could be related to Alfv\'{e}n waves excited near the surface of a neutron star \cite{Israel_1, Wang_1}. Energetic starquake drives Alfv\'{e}n waves in the magnetosphere of the source object, and a part of dissipated energy of the waves accelerate particles \cite{Blaes_1, Holcomb_1}. Chen et al. \cite{Chen_1} proposed an idea of Alfv\'{e}nic wakefield acceleration near a relativistic shock. The intense Alfv\'{e}n waves are thought to be excited also in an accretion disk of a black hole. Ebisuzaki and Tajima \cite{Ebisuzaki_1, Tajima_1, Ebisuzaki_2} discussed the scenario that the Alfv\'{e}n waves are converted to the electromagnetic (EM) waves as they propagate in the jet along the field line in a rarefied plasma, and particles are energized through the wakefield acceleration. Chang et al. \cite{Chang_1} proposed the theory of relativistic magnetowave-induced plasma wakefield acceleration (MPWA) and demonstrated, through numerical simulations, the excitation of wakefields driven by right-hand circularly polarized Alfv\'{e}n (whistler mode) waves. However, the generation of relativistic particles by MPWA was not confirmed in the simulation. 
%In the non-relativistic regime, Kuramitsu and Krasnoselskikh~\cite{Kuramitsu_1} proposed gyroresonant surfing acceleration driven by the combined action of an electrostatic potential and circularly polarized electromagnetic waves. In this mechanism, an electrostatic potential, like that observed near quasi-parallel shocks \cite{Behlke_2003}, forces particles to maintain a resonant condition.

%In MPWA, the strength of the ambient magnetic field, i.e., the group velocity of the Alfvén waves, must be sufficiently large to maintain long acceleration length. However, one difficulty arises with the ponderomotive force when particles are strongly magnetized:
%\begin{equation}
%f_{ps}=-\frac{1}{2}\frac{m_{s}c^{2}}{\gamma_{s}}\frac{\omega}{\omega-\omega_{cs}/\gamma_{s}},
%\label{eq1}
%\end{equation}
%where c, $\omega$, $m_{s}$, $\omega_{cs}$, $\gamma_{s}$ are speed of light, wave frequency, mass, gyro frequency and relativistic factor of $s$th species. When particles are initially non-relativistic, the ponderomotive force becomes small, and the wakefield is hardly excited compared to the non-magnetized case.

In our previous study \cite{Isayama_1}, we demonstrated the relativistic particle acceleration in counter-propagating Alfv\'en waves, where efficient energy conversion from waves to particles occurs when their amplitudes exceed critical values. This substantial increase of energy conversion efficiency is associated with phase transitions in the behavior of particles trapped within a magnetic envelope trough. Above these critical amplitudes, any particles irreversibly gain relativistic energy via relativistic cyclotron resonance within a short time, regardless of their initial position and energy. Such counter-propagating Alfv\'en waves are naturally generated in the course of preceding successive decay instabilities. Matsukiyo \& Hada \cite{Matsukiyo_1} showed that a relativistic Alfv\'en wave in an electron-positron plasma is unstable to form locally enhanced counter-propagating Alfv\'en waves, leading to efficient particle acceleration. 

In this study, we propose a particle acceleration mechanism driven by a large-amplitude Alfv\'en wave packet in a strong magnetic field, accompanied by self-generated counter-propagating waves via decay instability. Once counter-propagating waves are generated, these waves resonantly accelerate particles perpendicular to the ambient magnetic field through counter-propagating wave-particle resonant acceleration (CWRA) \cite{Isayama_1}. Efficient particle acceleration persists even in highly asymmetric cases where the amplitudes of counter-propagating waves differ significantly, provided that the parent wave amplitude exceeds the critical threshold and magnetic envelope troughs form, enabling particle trapping. The bulk electron energy subsequently increases, accompanied by a decrease in the effective electron cyclotron frequency ($\omega_{ce}/\gamma_{e}$). Consequently, the dispersion effect of the whistler-mode branch leads to the onset of modulational instability. This electrostatic field enhances resonant acceleration through gyroresonant surfing acceleration (GRSA) \cite{Kuramitsu_1}, while the $\mathbf{v} \times \mathbf{B}$ force converts perpendicular momentum into parallel momentum along the phase-space trajectory determined by the resonance condition in terms of the parent wave. In this process, modulational electrostatic fields help to keep the resonance condition. Even after GRSA ceases due to the  diminishing electrostatic field acting on the particles, these particles become trapped by the parent wave and undergo continuous acceleration through single-wave resonant acceleration (SWRA). %This sequential acceleration process begins with CWRA, followed by energy enhancement through GSA. Finally, WTA sustains the continuous acceleration of particles to highly relativistic energies.

%Additionally, the accumulation of accelerated particles induces electrostatic trailing fields behind and at the tail of the wave packets. This process leads to trailing-field acceleration (TFA) of particles in the tail of the Alfv\'en wave packet. Unlike conventional wakefield excitation, which primarily relies on the wave envelope gradient, RTFA is predominantly driven by a bunch of resonantly accelerated electrons, with additional support from the ponderomotive force of the wave packet.

%The multi-stage acceleration process, consisting of CWRA, GRSA, followed by WTA, results in electron energies that significantly exceed those achieved by TFA alone. These mechanisms are initiated when the Alfv\'en wave amplitude surpasses the CWRA threshold \cite{Isayama_1}. Furthermore, the phase-space trajectories of accelerated electrons provide valuable insights into the underlying acceleration processes.

To assess the effectiveness of the proposed multi-stage particle acceleration mechanism, a one-dimensional particle-in-cell (PIC) simulation is performed in Section II. Section III presents a detailed analysis of electron trajectories in a monochromatic wave, providing deeper insights into the dynamics of resonant acceleration and particle wave trapping. Finally, Section IV summarizes the key findings of this study and explores the broader implications of the proposed acceleration mechanism in astrophysical contexts.

 %These particles are trapped in the propagating wave packets and continuously accelerated in a manner similar to gyroresonant surfing acceleration \cite{Kuramitsu_1}. While non-relativistic gyroresonant acceleration requires an electrostatic potential to maintain the particle's resonant condition, particles in this process satisfy the resonant condition through self-generated counter-propagating waves. 

%Once particles are resonantly accelerated by this mechanism, the denominator of (\ref{eq1}) becomes sufficiently small, and the particles experience a strong ponderomotive force, leading to the excitation of wakefields. In this Letter, we demonstrate that relativistic wakefield acceleration becomes possible when the wave amplitude of the Alfv\'en wave packet exceeds a critical value. Furthermore, we show that the relativistic resonant particle acceleration is the dominant acceleration process in a propagating wave packets. 
\begin{figure*}[t]
 \centering
 \includegraphics[clip, width=2.0\columnwidth]{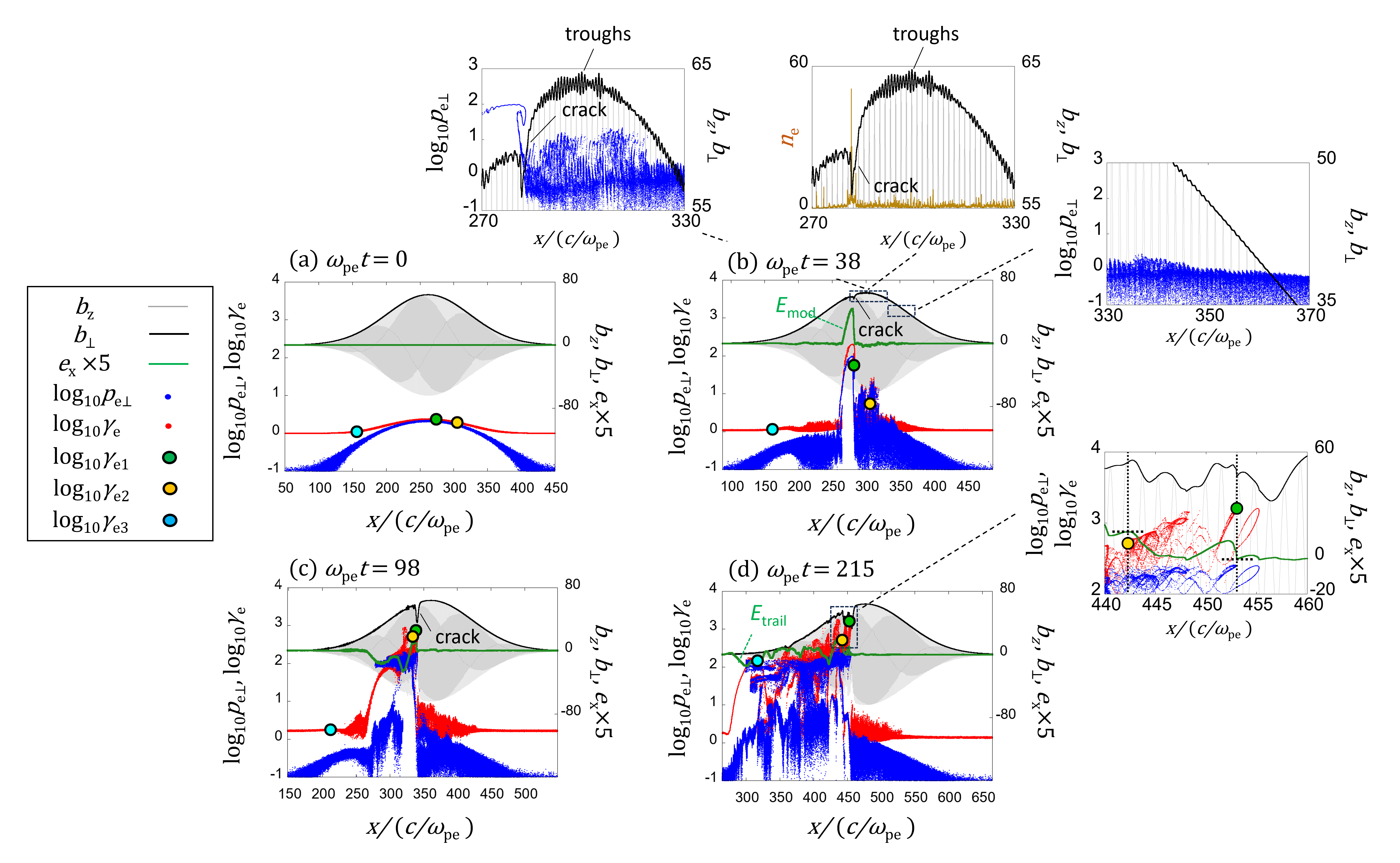}% Here is how to import EPS art
 \caption{Results when the wave amplitude is supercritical $b_{w}=B_{w}/B_{0}=0.8$, where $B_{0}$ is the strength of the background magnetic field. Spatial profiles of the normalized wave magnetic field $b_{z}=eB_{z}/m_{e}\omega_{pe}$ (gray), the wave envelop $b_{\perp}=\sqrt{b_{y}^{2}+b_{z}^{2}}$ (black), the electrostatic field $e_{x}\times5=eE_{x}/m_{e}c\omega_{pe}\times5$ (magenta) and electron's perpendicular momentum ($p_{e\perp}$: blue dots) and energy ($\gamma_{e}$: red dots) are shown at (a) $\omega_{pe}t=0$, (b) $\omega_{pe}t=38$, (c) $\omega_{pe}t=98$ and (d) $\omega_{pe}t=215$. The large amplitude modulational electrostatic field ($E_{\text{mod}}$)} and the induced trailing field ($E_{\text{trail}}$) are indicated in panels (b), (c) and (d), respectively. Three particles are represented by different colors for their energies: green ($\gamma_{e1}$), yellow ($\gamma_{e2}$), and cyan ($\gamma_{e3}$). The embedded figures in (b) provide enlarged views of the electron density profile, the transverse electron momentum $p_{e\perp}$, and the magnetic envelope. The embedded figure in (d) provides an enlarged view of the same profiles as in the original figure, where the vertical and horizontal black dotted lines represent the positions of the green and yellow electrons and the intensity of $e_{x}$ at these positions, respectively.
 \label{fig1}
\end{figure*}%

\begin{figure}[t]
 \centering
 \includegraphics[clip, width=0.8\columnwidth]{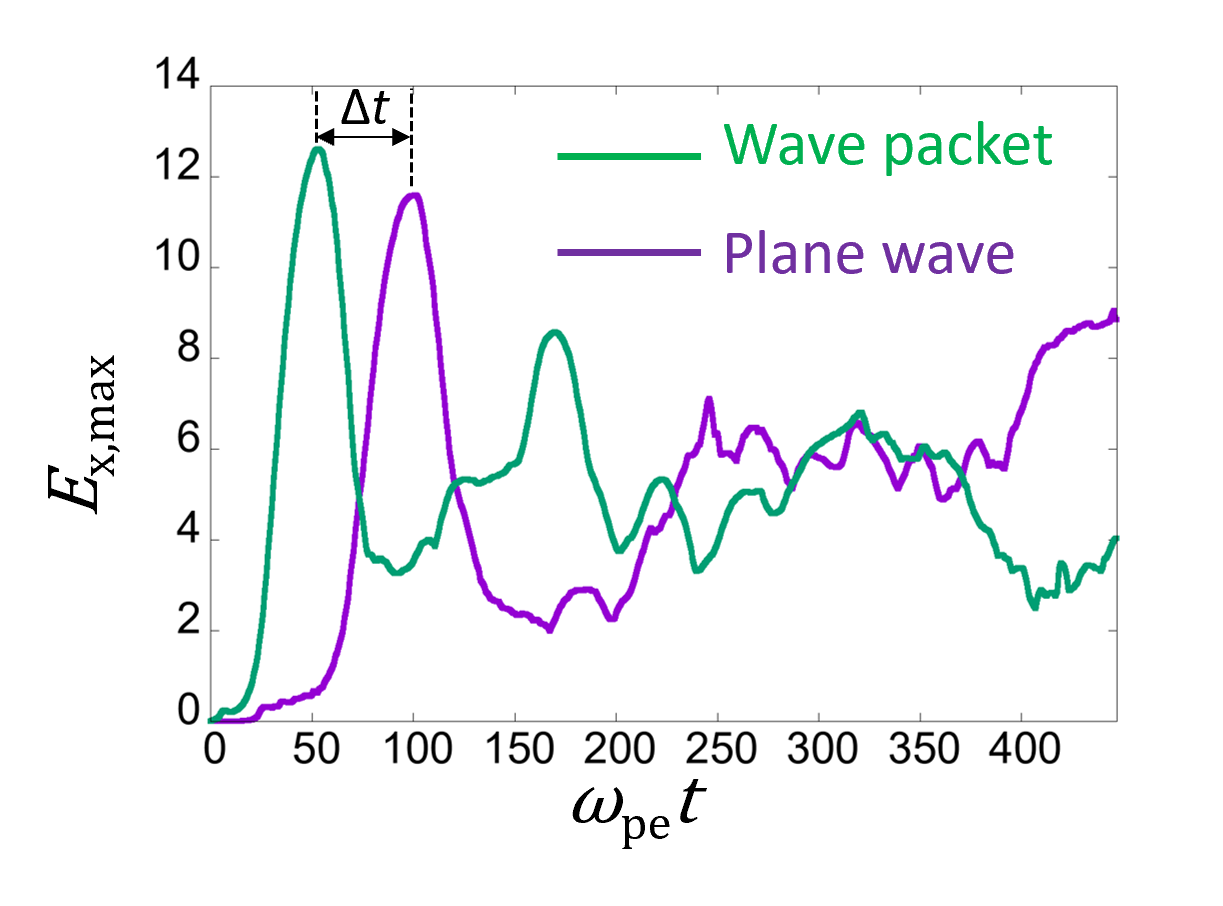}% Here is how to import EPS art
 \caption{Temporal evolution of $E_{x,\text{max}}$ for the cases of a Gaussian envelope and a plane wave.}
 \label{fig2}
\end{figure}%

\begin{figure*}[t]
 \centering
 \includegraphics[clip, width=2.0\columnwidth]{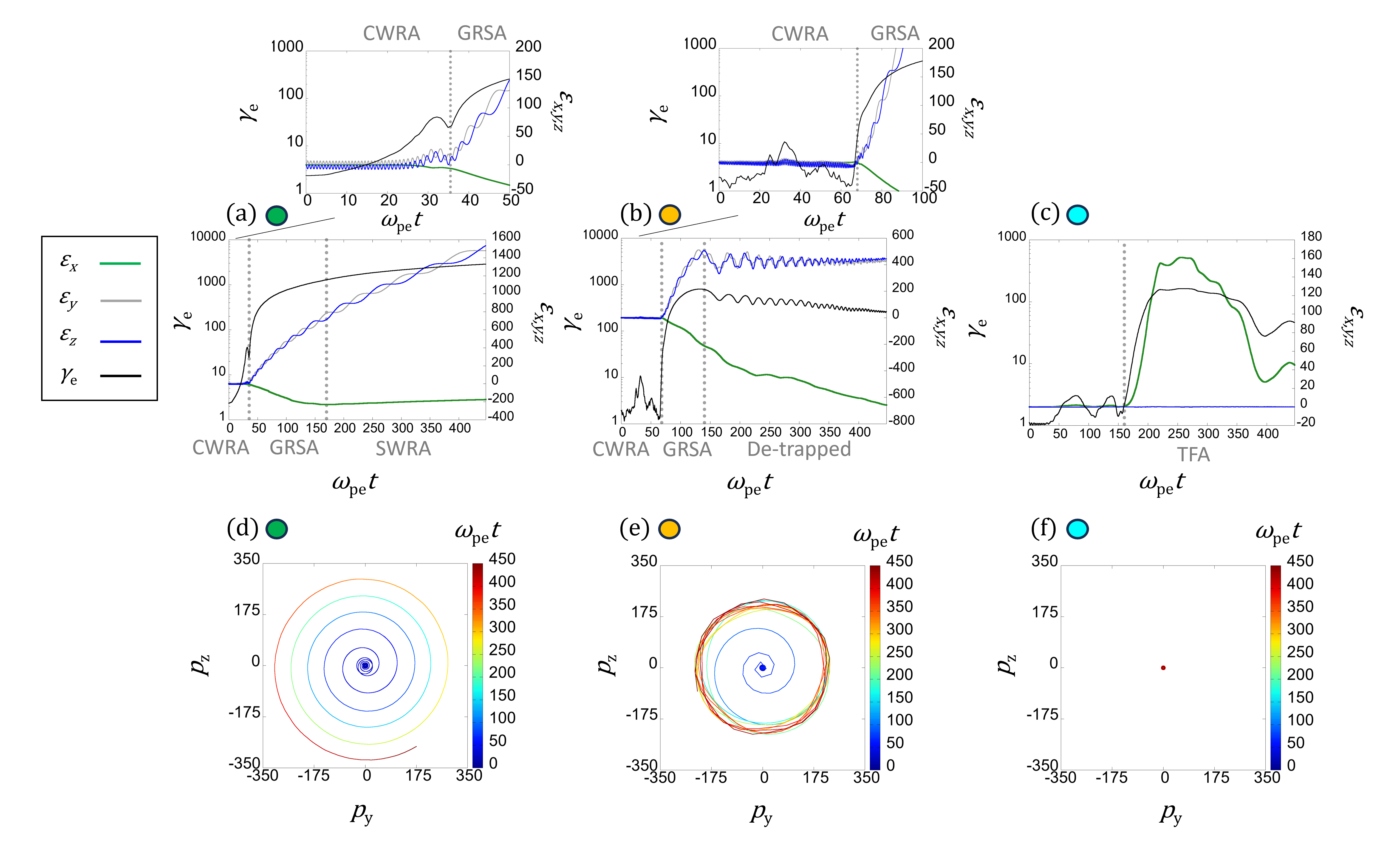}% Here is how to import EPS art
 \caption{Temporal profiles of the energy gain from the $E_{x}$ ($\epsilon_{x}$) (green), $E_{y}$ ($\epsilon_{y}$) (gray), and $E_{z}$ ($\epsilon_{z}$) (blue) components, along with the total energy $\gamma_{e}$ (black) with logarithmic scale, for the three colored electrons; (a) green, (b) yellow, and (c) cyan, corresponding to Fig.~\ref{fig1}. Each time region separated by the gray dotted line indicates a different acceleration process, namely CWRA, GRSA, SWRA, and TFA. The bottom panels display the trajectories of the (d) green, (e) yellow, and (f) cyan electrons in the $p_{y}-p_{z}$ momentum space, represented by colored lines with a time scale.}
 \label{fig3}
\end{figure*}%
\begin{figure*}[t]
 \centering
 \includegraphics[clip, width=2.0\columnwidth]{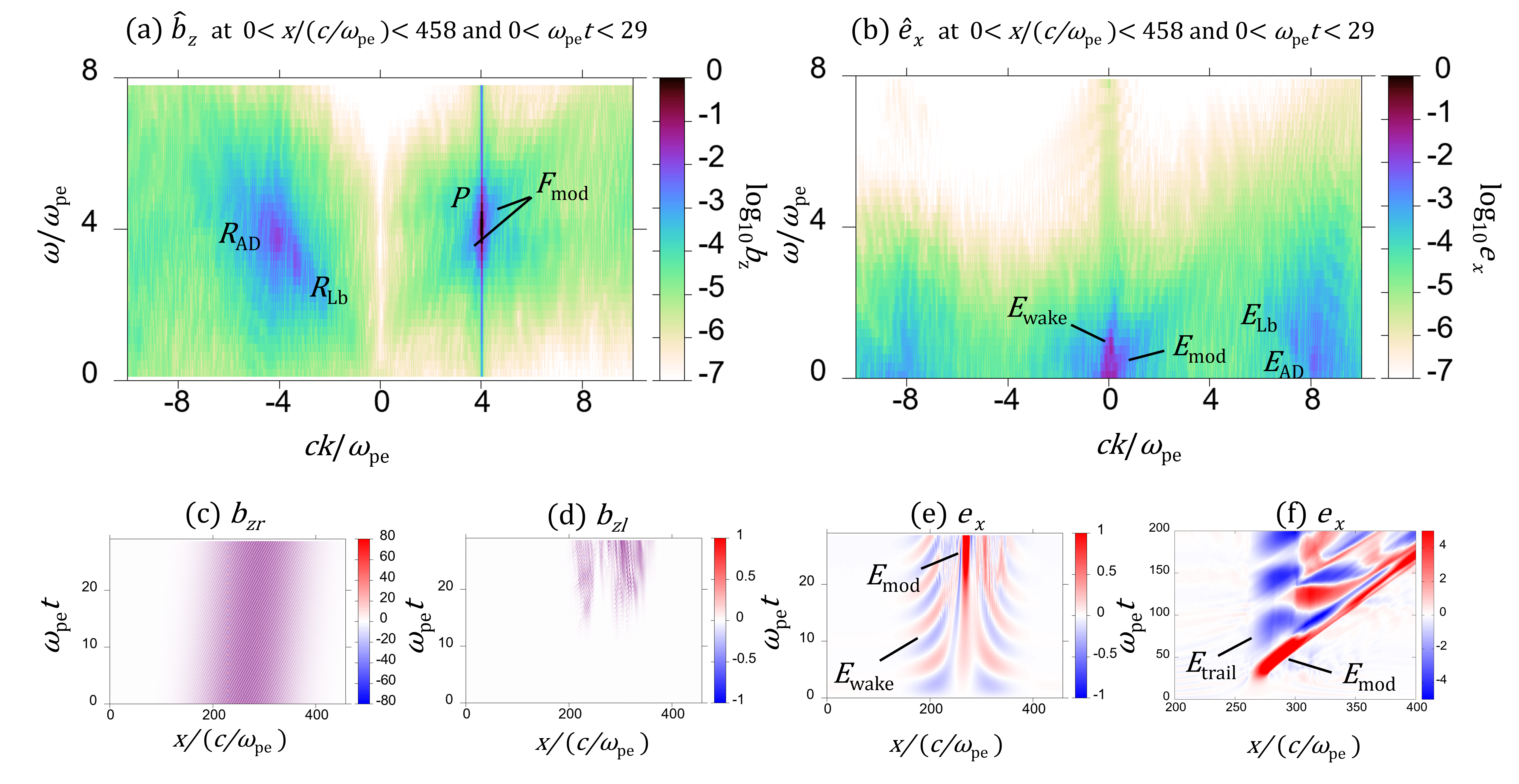}% Here is how to import EPS art
 \caption{The wave power spectra for (a) the \(\hat{b}_{z}\) component and (b) the \(\hat{e}_{x}\) component. The spectra in (a) reveal the parent wave (P) and two antiparallel propagating electromagnetic waves (\( R_{\text{AD}} \) and \( R_{\text{Lb}} \)), and modulational wave near the parent wave ($F_{\text{mod}}$)}. The spectra in (b) exhibit two parallel propagating electrostatic waves (\( E_{\text{AD}} \) and \( E_{\text{Lb}} \)), as well as wakefield (\( E_{\text{wake}} \)) and modulational field (\( E_{\text{mod}} \)). The corresponding temporal and spatial profiles of (c) the right-hand helicity mode ($b_{zr}$), (d) the left-hand helicity mode ($b_{zl}$), and (e) the electric field component $e_{x}$ during the early stage ($\omega_{pe}t = 0\text{--}57$) over the spatial domain $0 < x / (c / \omega_{pe}) < 458$ are presented. (f) The evolution of \( e_x \) over a longer time interval (\( \omega_{pe}t = 0\text{--}200 \)) within the range \( 200 < x / (c / \omega_{pe}) < 400 \), highlighting the presence of the trailing field (\( E_{\text{trail}} \)) behind \( E_{\text{mod}} \).
 \label{fig4}
\end{figure*}%

\section{1D PIC SIMULATION}
In this simulation, long time evolution of a wave packet of a right-hand circularly polarized wave propagating along a strong ambient magnetic field $B_{0}$ in the $x$ direction is investigated. The system size of the periodic domain is set to $L_{x}/(c/\omega_{pe}) = 1832$, which is sufficiently large to prevent negatively propagating waves from affecting the results. To ensure that the group velocity of the wave packet approximates the speed of light ($v_{g}\sim c$), the magnetization parameter of the electron is set to $\sigma_{0}=\Omega^{2}_{ce}/\omega^{2}_{pe}=6.4\times10^{3}$. The wavenumber and the frequency of the carrier wave are set as $k\sim4.0\ \omega_{pe}/c$ and $\omega\sim3.99\ \omega_{pe}$, respectively, according to the dispersion relation \cite{Matsukiyo_2}

\begin{equation}
    \frac{c^2k^2}{\omega^2}=1-\sum_{j}\frac{\omega_{pj}^2}{\omega(\gamma_{j}\omega+\Omega_{cj})}
\label{eq1}
\end{equation}
where $\omega_{pj}$, $\Omega_{cj}$, and $\gamma_{j}=1/\sqrt{1-v_{j}^2/c^2}$ are representing the plasma frequency, the non-relativistic cyclotron frequency, and the Lorentz factor of $j$th species. The width of Gaussian wave packet is $w_{0}/\Delta x=5.0/\sqrt{2}\times10^{4}$, where the grid size $\Delta x$ is set as $1.7\times10^{-3}\ c/\omega_{\text{pe}}$. The time step size is sufficiently small to resolve the gyro motion of non-relativistic electrons, i.e., $\Delta t\Omega_{\text{ce}}=0.14$. The transverse bulk velocities of electrons and ions in the packet are estimated using the Walen relation \cite{Hollweg_1}, with the real mass ($m_{i}=1836\ m_{e}$) of ions, while the particles have zero bulk velocity in the $x-$direction. The initial temperature of electrons and ions are set to $T_{e}/m_{e}c^2=1.0\times10^{-3}$ and $T_{i}=T_{e}m_{e}/m_{i}$, respectively.

The initial spatial profiles of fields, electron momentum, and energy are shown in Fig.~\ref{fig1}(a). In the early stage at $\omega_{pe} t = 38$ (Fig.~\ref{fig1}(b)), as shown in the enlarged view, small troughs of magnetic envelope are created in the range $270<x/(c/\omega_{pe})<330$, where counter-propagating waves are excited via decay instability. In this region, electrons are accelerated perpendicular to the ambient magnetic field through CWRA at the troughs of the magnetic envelope. The large crack of the Alfvén wave packet at $x/(c/\omega_{pe}) \sim 280$ arises from modulational instability associated with the localized electrostatic field $E_{\text{mod}}>0$. The ponderomotive force of the Alfvén wave packet facilitates the formation of electron bunches and thereby locally amplifying $E_{\text{mod}}$. Once it exceeds a certain threshold, the field assists resonant acceleration via gyroresonant surfing acceleration (GRSA), which further accumulates electron bunches and leads to the subsequent formation of the crack structure. In later time at $\omega_{pe} t = 98$ and 215 (Figs.~\ref{fig1}(c) and (d)), electrons are further accelerated. At $\omega_{pe}t=215$, electrostatic trailing fields, $E_{\text{trail}}$, has been generated behind the modulational electrostatic field, $E_{\text{mod}}$ (Fig.~\ref{fig1}(d)).

The ponderomotive force locally amplifies $E_{\text{mod}}$, as evident in the temporal evolution of the maximum electrostatic field ($E_{x,\text{max}}$) shown in Fig.~\ref{fig2}. The purple line represents $E_{x,\text{max}}$ in the case of homogeneous wave amplitude (without Gaussian wave packet) for comparison. For the homogeneous case, the system size of the periodic domain is set to $L_{x}/(c/\omega_{pe}) = 458$, which is comparable to the size of the Gaussian wave packet. All other parameters remain the same as those in the Gaussian wave packet case. As shown in Fig.~\ref{fig2}, the growth of $E_{x,\text{max}}$ in the Gaussian wave packet case (green line) is faster than that in the homogeneous case (purple line) by $\omega_{pe} \Delta t \sim 47$. 

%\textcolor{red}{Before discussing the acceleration process, we first define the key terminology used in this study. The wave packet refers to a wave structure characterized by a Gaussian profile. The troughs are defined as modulated structures of the wave packet created by counter-propagating waves. The crack is a strongly modulated region of the wave packet resulting from the accumulation of an accelerated electron bunch. The charge separation field $E_{\text{cs}}$ is the electrostatic field formed between the electron bunch and the background immobile ions. The trailing field $E_{\text{trail}}$ is the electrostatic field excited behind and at the tail of the wave packet.}

%It is observed that when $E_{x} > 0$ is excited within the wave packet at $\omega_{pe}t = 47$ (Fig.~\ref{fig1}(b)), the resonant acceleration of electrons is enhanced, similar to GSA \cite{Kuramitsu_1}, and further accumulates the electron bunch. This positive feedback promotes the formation of localized electron bunches.
%Simultaneously, the perpendicular momentum is converted into parallel momentum via the $\mathbf{v} \times \mathbf{B}$ force. As will be demonstrated later, these electrons follow the phase-space trajectory determined by the resonance condition and the wave trapping condition, which subsequently converts $p_{e\perp}$ into positive $p_{ex}$. Additionally, $E_{\text{cs}}$ drives the generation of electrostatic trailing fields, $E_{\text{trail}}$, behind and at the tail of the wave packet at $\omega_{pe}\Delta t = 215$ (Fig.~\ref{fig1}(d)).

To address the acceleration process, Fig.~\ref{fig1} highlights three electrons in different colors, each representing an example of electrons undergoing different acceleration mechanisms. 

The green electron, initially located at $x/(c/\omega_{pe})\approx274$ (Fig.~\ref{fig1}(a)), is first accelerated by CWRA. It is further accelerated when the modulational electrostatic field $E_{\text{mod}} (> 0)$ is created at $\omega_{pe} t = 47$ and 98 (Figs.~\ref{fig1}(b) and (c)). As shown later, this acceleration phase corresponds to GRSA. At $\omega_{pe} t = 215$, as shown in the enlarged view in Fig.~\ref{fig1}(d), the electrostatic field experienced by this particle diminishes ($E_{x} \sim 0$), leading to the termination of GRSA. Subsequently, the particle is continuously accelerated while being trapped by the parent wave through SWRA.

In a similar process, the yellow electron, initially located at $x/(c/\omega_{pe}) \approx 305$ (Fig.~\ref{fig1}(a)), is first accelerated by CWRA (Fig.~\ref{fig1}(b)), and is then further accelerated by GRSA (Fig.~\ref{fig1}(c)). However, the electron energy is almost unchanged after that ($\omega_{pe}t =215$) as seen in Fig.~\ref{fig1}(d), since $E_x (>0)$ is too strong to sustain CRSA as explained later.

The cyan electron, initially located at the tail of the packet at $x/(c/\omega_{pe}) \approx 156$ (Fig.~\ref{fig1}(a)), is accelerated only in the later time ($\omega_{pe}t=215$, Fig.~\ref{fig1}(d)) by the trailing field $E_{\text{trail}}(<0)$. This process is then termed trailing-field acceleration (TFA).

Figure~\ref{fig3} illustrates the time evolution of the energy gain from each component of the electric field for these three electrons. 

$$
\varepsilon_{x,y,z} = -{e \over m_e c^2} \int^t_0 E_{x,y,z}v_{x,y,z} dt
$$

The acceleration of the green electron is initiated by CWRA, reaching $\gamma_{e} \sim 20$ while $E_{x} \sim 0$ for $\omega_{pe}t < 36$ (Figs.~\ref{fig3}(a) and (b)). After that, resonant acceleration by $E_{y}$ and $E_{z}$ occurs via GRSA. In this time domain ($36 < \omega_{pe}t < 170$), the $-e{\bf v} \times {\bf B}$ force acting in the $x-$direction is partially canceled by the force $-e E_x$, allowing the electron to remain in resonance with the parent wave while maintaining a constant relative phase over a long period. This can be confirmed from Fig. 3(d), where the particle trajectory in $p_{ey}-p_{ez}$ space maintains a spiral shape while its momentum increases monotonically. During GRSA, their perpendicular momentum is converted into parallel momentum via the $\bf{v}\times\bf{B}$ force, as indicated by $\gamma_{e} >> p_{e,\perp}$.

For $\omega_{pe} t > 170$, the contribution from $E_{x}$ exhibits a slightly increasing trend (Fig.~\ref{fig3}(a)) due to the emergence of a slightly negative $E_{x}$ field, consequently leading to the termination of GRSA. Nonetheless, acceleration persists via SWRA, enabling $\gamma_{e}$ to reach $\sim 2840$ at $\omega_{pe} t = 447$. Similarly, the acceleration of the yellow electron is initiated by CWRA, reaching $\gamma_{e} \sim 10$ while $E_{x} \sim 0$ for $\omega_{pe} t < 68$, and resonant acceleration is enhanced via GRSA at $\omega_{pe} t > 68$ (Figs.~\ref{fig3}(b) and (e)). However, for $\omega_{pe} t > 140$, $\gamma_{e}$ exhibits a decreasing trend. This decline is attributed to the presence of a strong, excessive electrostatic field, $E_{x} > 0$, as shown in the enlarged view of Fig.~\ref{fig1}(d), which is too strong to sustain GRSA. For the cyan electron (Fig.~\ref{fig1}(c)), acceleration does not occur until the energy gain from $E_{\text{trail}} < 0$ becomes significant at $\omega_{pe} t > 160$, where it is accelerated solely by TFA, without any resonant acceleration in the $p_{ey}-p_{ez}$ space (Fig.~\ref{fig1}(f)).

\begin{figure}[t]
 \centering
 \includegraphics[clip, width=0.8\columnwidth]{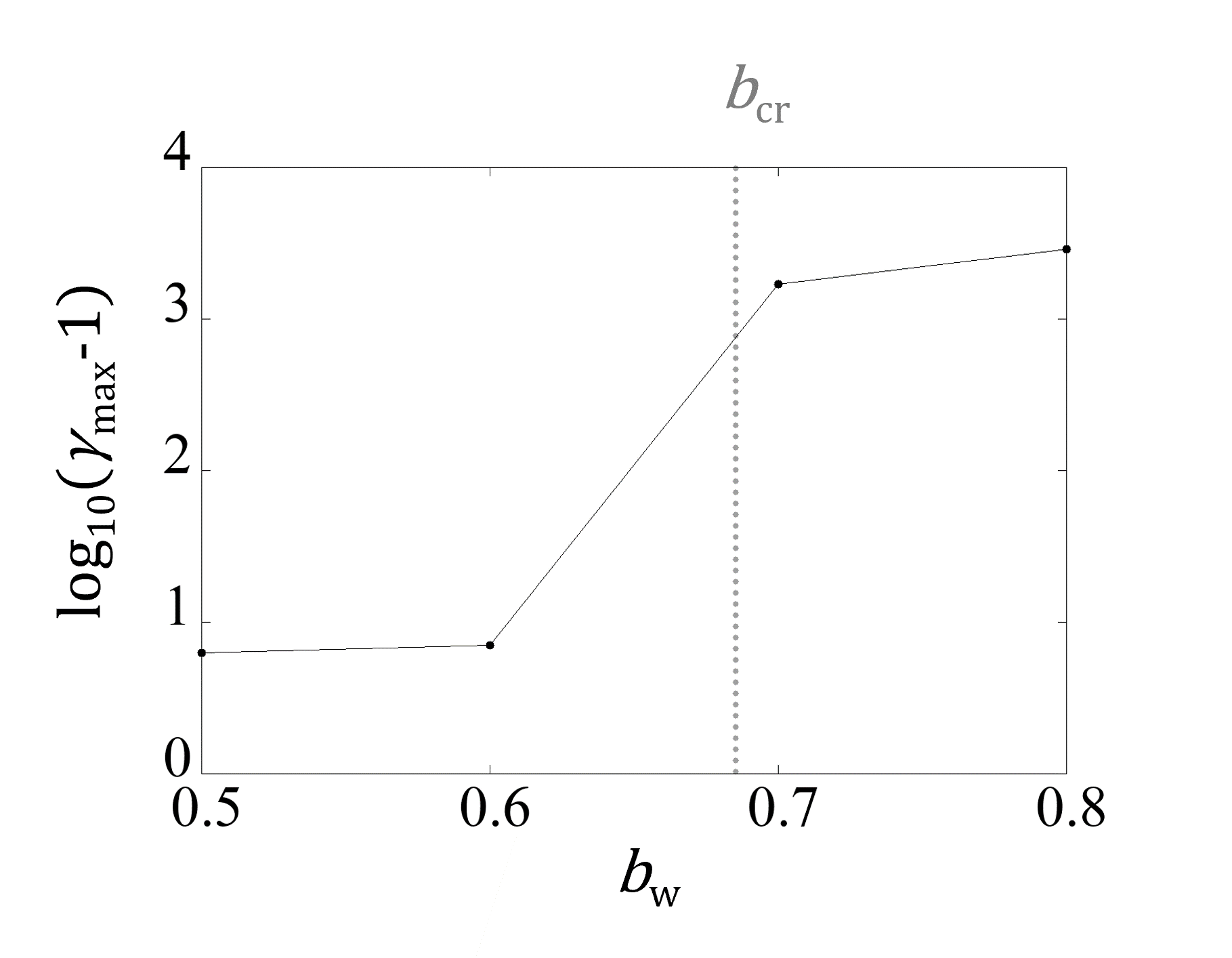}% Here is how to import EPS art
 \caption{The depdendence of the maximum electron energy on the wave amplitude at $\omega_{pe}t=447$. The critical amplitude $b_{\text{cr}}=0.685$ is estimated by $(1-\sqrt{2\nu})/v_{\text{ph}}$ \cite{Isayama_1}.}
 \label{fig5}
\end{figure}%

The wave amplitude spectra of the $b_{z}$ and $e_{x}$ components are presented in Figs.~\ref{fig4}(a) and (b). The right-hand ($b_{zr}$) and the left-hand ($b_{zl}$) helicity mode, corresponding to positive and negative wavenumbers respectively, are decomposed using the Fourier decomposition technique \cite{Terasawa_1}. As discussed by Matsukiyo and Hada \cite{Matsukiyo_2}, the parent wave ($P$) decays into two antiparallel propagating electromagnetic waves ($R_{\text{AD}}$ and $R_{\text{Lb}}$) and two parallel propagating electrostatic waves ($E_{\text{AD}}$ and $E_{\text{Lb}}$). The waves $R_{\text{AD}}$ and $E_{\text{AD}}$ are generated by acoustic decay instability, while $R_{\text{Lb}}$ and $E_{\text{Lb}}$ arise from Langmuir decay instability. The daughter waves $F_{\text{mod}}$ in $\hat{b}_{z}$ near the parent wave (P) (Fig.~\ref{fig4}(a)) and $E_{\text{mod}}$ in $\hat{e}_{x}$ (Fig.~\ref{fig4}(b)) are also observed. The spectral feature of these waves ($F_{\text{mod}}$ and $E_{\text{mod}}$) indicate that modulational instability occurs in this stage, during which the bulk electron energy ($\gamma_e$) increases, as seen in Fig.~\ref{fig1}(b), accompanied by a decrease in the effective electron cyclotron frequency ($\omega_{ce}/\gamma_{e}$). As a result, the dispersion effect of whistler mode branch becomes non-negligible so that the modulational instability sets in.  The electrostatic wave $E_{\text{wake}}$, identified as the small amplitude wakefield, oscillates around $\omega/\omega_{pe} = 1$ and is driven by the ponderomotive force of the Alfv\'en wave packet, which is the wakefield discussed in \cite{Chang_1}. The modulational electrostatic field $E_{\text{mod}}$ remains in a growth phase at this time, as illustrated in Fig. \ref{fig2} and Fig. \ref{fig4} (e). The ponderomotive force exerted on electrons with negligible parallel momentum ($p_{e\parallel}\sim0$) in a magnetized plasma is expressed as follows \cite{Sharma_1}:
\begin{equation}
f_{pe}=-\frac{1}{2}\frac{m_{e}c^{2}}{\gamma_{e}}\frac{\omega}{\omega-\Omega_{ce}/\gamma_{e}}\frac{\partial a_{0}^{2}}{\partial x},
\label{eq2}
\end{equation}
where $a_{0}=eE_{\perp}/m_{e}c\omega$ is the strength parameter of the Alfv\'en wave. When particles are initially non-relativistic, the ponderomotive force is relatively weak, particularly in strongly magnetized environments, making it difficult to excite large-amplitude wakefields. Alternatively, resonant acceleration and the parallel transport of electrons via the $\mathbf{v} \times \mathbf{B}$ force amplify the modulational electrostatic field ($E_{\text{mod}}$) (Fig.~\ref{fig4} (e) and (f)), which subsequently drive electrostatic trailing-fields ($E_{\text{trail}}$) extending behind $E_{\text{mod}}$ (Fig.~\ref{fig4} (f)). As shown in Fig.~\ref{fig2}, the ponderomotive force facilitates the formation of electron bunch concentration during the CWRA by the attracting force of $f_{pe} \propto \partial a_{0}^{2}/\partial x$, which arises due to the negative denominator in (2), until $\gamma_{e}$ reaches $\Omega_{ce}/\omega = 20$. This, in turn, promotes the amplification of the modulational electrostatic fields ($E_{\text{mod}}$) in the early phase at $\omega_{pe}t = 47$ (Fig.~\ref{fig1}(b) and Fig. 2). Unlike conventional wakefield acceleration, which primarily relies on the wave packet gradient, TFA is predominantly driven by a bunch of resonantly accelerated electrons.

\begin{figure*}[t]
 \centering
 \includegraphics[clip, width=2.0\columnwidth]{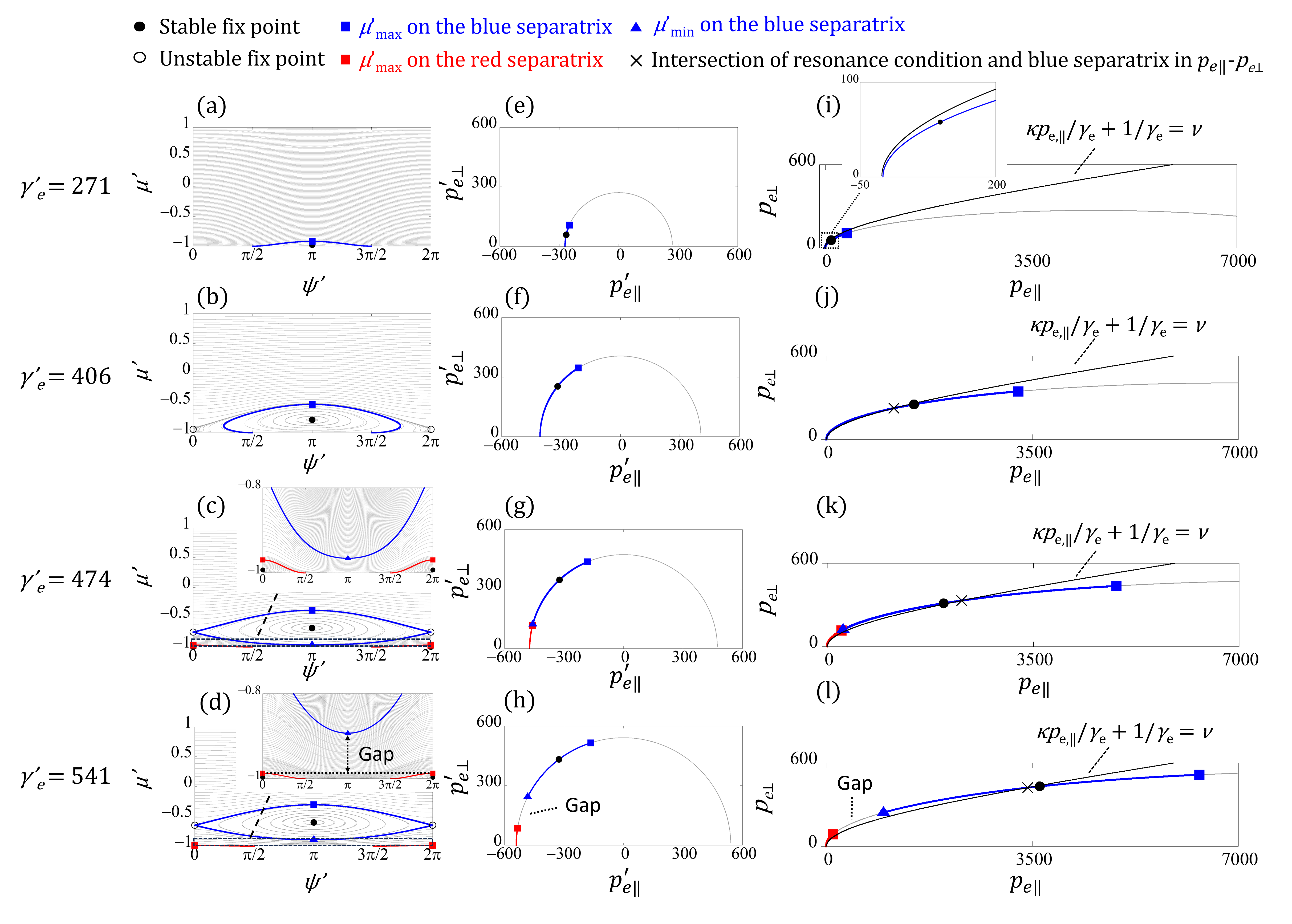}% Here is how to import EPS art
 \caption{Electron trajectories with fixed energy in the wave frame for four different values of $\gamma'_{e}=271, 406, 474$, and 541. (a)-(d) The $\psi'$-$\mu'$ space. The stable fixed points are marked by black solid circles, while unstable fixed points are marked by black hollow circles. The trapping regions exist at $\mu' < 0$. The trapping regions centered at $\psi' = \pi$ and at $\psi' = 0, 2\pi$ are bounded by the blue and red separatrix, respectively. The maximum $\mu'_{\text{max}}$ on the separatrix are denoted by blue and red solid squares. The minimum $\mu'_{\text{min}}$ on the blue separatrix are denoted by blue solid triangles in (c) and (d). (e)–(h) The $p'_{e,\parallel}$-$p'_{e,\perp}$ momentum space in the wave frame. (i)–(l) The $p_{e,\parallel}$-$u_{e,\perp}$ momentum space in the simulation frame. The black line represents the resonance condition, \(\kappa p_{e,\parallel}/\gamma_{e} + 1/\gamma_{e} = \nu\), for electrons. The intersection of the resonance condition and the trajectory on the blue separatrix is represented by black cross. In (e)–(l), the trajectory on the blue and red separatrix, the stable fixed points (black solid circles), as well as the maximum \(\mu'_{\text{max}}\) (blue and red solid squares) and minimum \(\mu'_{\text{min}}\) (blue solid triangles) on the separatrix are also projected. In (d), where the fixed energy is set to \(\gamma'_{e} = 541\), a gap emerges between the minimum \(\mu'_{\text{min}}\) (blue solid triangle) on the blue separatrix and the maximum \(\mu'_{\text{max}}\) (red solid squares) on the red separatrix. This gap is also projected in (h) and (i).}
 \label{fig6}
\end{figure*}%
\begin{figure*}[t]
 \centering
 \includegraphics[clip, width=2.0\columnwidth]{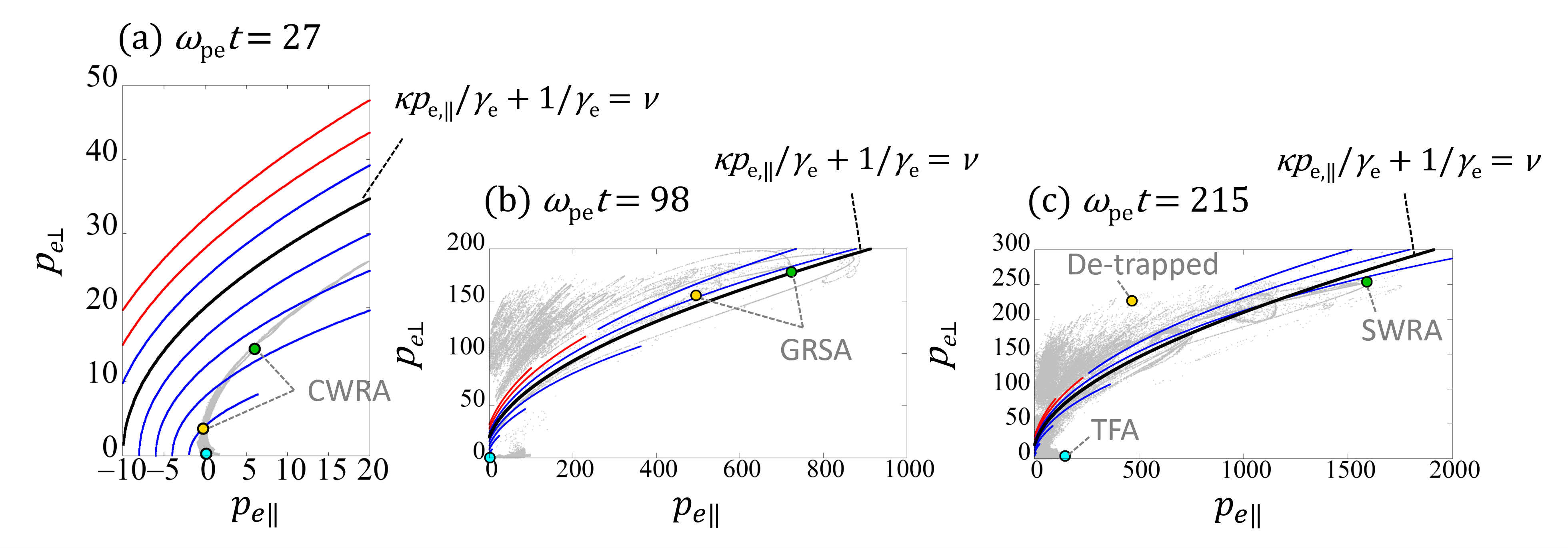}% Here is how to import EPS art
\caption{Electron trajectories are plotted as gray dots in the \( p_{e,\parallel} \)-\( p_{e,\perp} \) space, obtained from the simulation at \( \omega_{pe}t = 27 \), 98, and 215. The colored electrons (green, yellow, and cyan) from Fig.~\ref{fig1} are also shown. Additionally, the trajectories on the blue and red separatrix are plotted for different fixed energy values corresponding to \( \gamma'_{e} = 70, 136, 204, 271, 338, 406, 473, \) and \( 541 \). The resonance condition, \( \kappa p_{e}/\gamma_{e} + 1/\gamma_{e} = \nu \), is represented by the black line.
 \label{fig7}}
\end{figure*}
 
The dependence of the maximum electron energy on the wave amplitude is shown in Fig. \ref{fig5}. When the wave amplitude exceeds the critical value, the electron energy exceeds 2000 due to relativistic cyclotron resonance via CWRA and GRSA, followed by SWRA, whereas it remains below 10 when the wave amplitude is below the critical value. Therefore, the critical amplitude estimated in the counter-propagating wave system \cite{Isayama_1} can also be applied to this process, suggesting that CWRA serves as the trigger for the above multi-stage acceleration.

\section{Analysis of electron trajectory in a monochromatic wave}

To understand the behavior of accelerated electrons after the CWRA, we analyze the motion of electrons in a monochromatic plane wave. In the wave frame, the magnetic field associated with the wave is given by
\begin{equation}
{\bm b}'_{w} \equiv \left(
\begin{array}{c}
b_{w}'^y \\
b_{w}'^z \\
\end{array}
\right)=b'_{w}\left(
\begin{array}{c}
\cos\kappa'\xi' \\
-\sin\kappa'\xi' \\
\end{array}
\right).
\label{eq3}
\end{equation}
Here, we used the normalized variables, $\xi=x\Omega_{ce}/c$, $\kappa=kc/\Omega_{ce}$, and $b_{w}=B_{w}/B_{0}$, where $\Omega_{ce}=|q_{e}|B_{0}/m_{e}$ is a non-relativistic gyro frequency of electrons. The prime notation (') denotes quantities in the wave frame. In this analysis, the wave amplitude, $b_w$, wave number, $\kappa$, and wave frequency, $\nu=\omega/\Omega_{ce}$, used are identical to those in PIC simulation (Fig. 1). The equation of motion is given by 
\begin{equation}
\frac{d\boldsymbol{p}'_{e}}{d\tau'}=-\frac{1}{\gamma_{e}'}\left[\boldsymbol{p}'_{e}\times\left(\bm{\hat{x}}+\boldsymbol{b}'_{w}\right)\right]
\label{eq4}
\end{equation}
with the normalized momentum $\bm{p}_{e}=\gamma_{e} \bm{v}_{e}/c$.
Since the wave electric field vanishes in the wave frame, the electron energy remains constant in this frame.
\begin{equation}
\gamma'_{e}=\sqrt{1+p'^{2}_{e}}
\label{eq5}
\end{equation}
By introducing $\bm{p}'_{e}=\left(p'_{e,\parallel}, p'_{e,\perp}\cos\phi', p'_{e,\perp}\sin\phi'\right)$, (\ref{eq4}) is written as follows.
\begin{equation}
\dot{p'}_{e,\parallel}=\frac{b'_{w}}{\gamma'_{e}}{p'}_{e,\perp}\sin\psi'
\label{eq6}
\end{equation}
\begin{equation}
\dot{p'}_{e,\perp}=\frac{b'_{w}}{\gamma'_{e}}{p'}_{e,\parallel}\sin\psi'
\label{eq7}
\end{equation}
\begin{equation}
\begin{split}
\dot{\psi'}=\frac{1}{\gamma'_{e}}+\kappa'\frac{p'_{e,\parallel}}{\gamma'_{e}}-\frac{p'_{e,\parallel}}{p'_{e,\perp}}\frac{b'_{w}}{\gamma'_{e}}\cos\psi'
\end{split}
\label{eq8}
\end{equation}
where $\psi'$ is the phase difference between the gyrophase of the electron and the electric field of the wave, i.e. $\psi'=\kappa'\xi'-\phi'$. Rewriting (\ref{eq6})-(\ref{eq8}) in terms of the pitch angle cosine $\mu'=p'_{e\parallel}/p'_{e}$ yields,
\begin{equation}
\dot{p'}_{e}=0
\label{eq9}
\end{equation}
\begin{equation}
\dot{\mu'}=\frac{b'_{w}}{\gamma'_{e}}\sqrt{1-\mu'^{2}}\sin\psi'
\label{eq10}
\end{equation}
\begin{equation}
\begin{split}
\dot{\psi'}=\frac{1}{\gamma'_{e}}+\kappa'\frac{p'_{e}\mu'}{\gamma'_{e}}-\frac{\mu'}{\sqrt{1-\mu'^{2}}}\frac{b'_{w}}{\gamma'_{e}}\cos\psi'
\end{split}
\label{eq11}
\end{equation}
This system has another constant of motion defined as

\begin{equation}
\begin{split}
H'(\psi',\mu')=\frac{p'_{e}\kappa'}{2\gamma'_{e}}\left(\mu'+\frac{1}{\kappa'p'_{e}}\right)^{2}-\frac{1}{2\kappa'p'_{e}\gamma'_{e}}\\
+\frac{b'_{w}}{\gamma'_{e}}\sqrt{1-\mu'^{2}}\text{cos}\psi'
\end{split} 
\tag{11}
\end{equation}
where $\dot{\mu'}=-\partial{H'}/\partial{\psi'}$ and $\dot{\psi'}=\partial{H'}/\partial{\mu'}$ are satisfied.
Here, four different values of fixed electron energy in the wave frame are considered to represent the characteristics of the electron trajectory in momentum space ($p'_{e,\parallel}$-$p'_{e,\perp}$ and $p_{e,\parallel}$-$p_{e,\perp}$) as well as ($\psi'$-$\mu'$) space. The electron trajectories are depicted in the $\psi'$-$\mu'$ space (Figs.~\ref{fig6}(a)--(d)), the $p'_{e,\parallel}$-$p'_{e,\perp}$ space (Figs.~\ref{fig6}(e)--(h)), and the $p_{e,\parallel}$-$p_{e,\perp}$ space (Figs.~\ref{fig6}(i)--(l)). When the fixed energy is small at $\gamma'_{e} = 271$, a small trapping region bounded by the blue separatrix exists around $\psi' = \pi$ and $\mu' \sim -1$ (Fig.~\ref{fig6}(a)). The stable fixed point (black solid circle) is located at the center of the trapping region at $\psi' = \pi$. As the fixed energy increases to $\gamma'_{e} = 406$, the trapping region expands, and unstable fixed points (black hollow circle) emerge at $\psi' = 0$ and $2\pi$ (Fig.~\ref{fig6}(b)). In Figs.~\ref{fig6}(e)-(l),electron trajectories (gray lines), stable fixed points (black solid circles), the trajectories on the blue and red separatrix, as well as the maximum \(\mu'_{\text{max}}\) (solid squares) and minimum \(\mu'_{\text{min}}\) (solid triangles) on the separatrix are projected onto the momentum space \(p'_{e,\parallel}\)-\(p'_{e,\perp}\) (Fig.~\ref{fig6}(e)--(h)) and \(p_{e,\parallel}\)-\(p_{e,\perp}\) (Fig.~\ref{fig6}(i)--(l)). The resonance condition for electrons (black line) is also shown in the \(p_{e,\parallel}\)-\(p_{e,\perp}\) space, where the fixed points (black solid circles) lie very close to the resonance points (black crosses), except for \(\gamma'_{e} = 271\) (Fig.~\ref{fig6}(i)) in which the resonance condition does not intersect with the trajectory on the gray line, as shown in the enlarged view in Fig.~\ref{fig6}(i). Yet, as shown in (8), the solution of \(\dot{\psi'} = 0\) slightly differs from the inear resonance condition, \(1/\gamma'_{e} + \kappa' p'_{e,\parallel} \mu' / \gamma'_{e} = 0\), in the wave frame with \(\nu' = 0\). This deviation arises due to the finite wave amplitude, which corresponds to the third term in (8). As the fixed energy further increases to $\gamma'_{e} = 474$ (Fig.~\ref{fig6}(c)), the bottom of the blue separatrix \(\mu'_{\text{min}}\) (blue solid triangle) detaches from $\mu' = -1$, and new trapping regions bounded by the red separatrix appear around $\psi' = 0$ and $2\pi$. Notably, as the fixed energy increases further to $\gamma'_{e} = 541$ (Fig.~\ref{fig6}(d)), the blue separatrix shifts toward larger $\mu'$ values, while the size of the red separatrix decreases. This leads to the formation of a gap between the bottom ($\mu'_{\text{min}}$) (blue solid triangle) of the blue separatrix and the top ($\mu'_{\text{max}}$) (red solid squares) of the red separatrix. Consequently, when electrons are located in this gap region in momentum space (Figs.~\ref{fig6}(h) and (l)), they become de-trapped and are unable to gain energy through cyclotron resonance. This gap region expands as the fixed energy increases further. In contrast, when electrons remain trapped, the maximum attainable momentum (blue solid squares) in the simulation frame increases as the electron energy \(\gamma'_e\) increases (Fig.~\ref{fig6}(i)--(l)).

In the simulation, the maximum momentum and energy of the initial electrons is given by \(p_{e\parallel}\sim 0\), \(p_{e\perp} \sim 2.3\), and \(\gamma_{e} \sim 2.5\), respectively, in the simulation frame. The initial pitch angle cosine in the wave frame is \(\mu' \sim -1\), with \(p'_{e\parallel} \sim -\gamma_{w} \gamma_{e}v_{ph} \sim -42.3\), \(p'_{e\perp} \sim 2.3\), and \(\psi' \sim \pi\), where $\gamma_w = 1/\sqrt{1-v^2_{ph}/c^2}$ and $v_{ph}=\nu/\kappa$. Consequently, in \(\psi'\)-\(\mu'\) space, electrons are accelerated from the bottom $(\psi', \mu') = (\pi, -1)$ to the top $\mu'_{\text{max}}$ (blue solid squares) of the blue separatrix. The maximum initial electron energy in the wave frame is given by \(\gamma'_{e} = \sqrt{1 + p'^{2}_{e\perp} + p'^{2}_{e\parallel}} \sim 42.4\). When there is no acceleration, the wave trapping region in \(\psi'\)-\(\mu'\) space is small, and the maximum attainable momentum in the wave frame and simulation frame is \((p'_{e\parallel}, p'_{e\perp}) \sim (-42.1, 4.6)\) and \((p_{e\parallel}, p_{e\perp}) \sim (3.1, 4.6)\), respectively. Once counter-propagating waves are excited and multi-stage acceleration begins, the accelerated electrons gain highly relativistic energy through CWRA, followed by GRSA. Consequently, their trajectories move into a larger wave trapping region in \(\psi'\)-\(\mu'\) space, leading to an increase in the maximum attainable energy in the simulation frame.

Figure~\ref{fig7} shows the electron trajectory obtained from the simulation, represented by gray dots, in the \( p_{e,\parallel} \)-\( p_{e,\perp} \) space for different acceleration phases. The trajectories on the blue and red separatrices for several fixed energy values of \( \gamma'_{e} \), as well as the resonance condition in the simulation frame, are also plotted. At the early stage of acceleration  ($\omega_{pe}t = 27$, Fig.~\ref{fig7}(a)), all the electrons including
 the green and yellow electrons are initially accelerated perpendicular to magnetic field. During this process, electrons do not follow the blue or red (i.e., gray) lines, indicating that they are accelerated via CWRA. Up to then, the green and yellow electrons are accelerated to $p_{e\perp} \sim 16$ and $p_{e\perp} \sim 3.6$. These two electrons are then further accelerated via GRSA. In this phase the electrons follow the black line, resonance condition, in Fig.~\ref{fig7}(b). This occurs with the help of the dragging force due to $E_{\text{mod}} (>0)$ forcing the electrons to stay close to the resonance condition. If the dragging force is too strong as for the yellow electron, it is detrapped from the resonance and stops being accelerated (Fig.~\ref{fig7}(c)). On the other hand, when $E_x$ felt by an electron becomes weak, GRSA terminates so that electron trajectory deviates from the black line and follows along the blue line (separatorix). This stage is the SWRA experienced by the green electron (Fig.~\ref{fig7}(c)).

When $E_{\text{mod}}$ becomes large enough to generate the trailing field, some electrons feel $E_x<0$, resulting in acceleration only in the $p_{e\parallel} > 0$ direction. This is the TFA, which is similar to the wakefield acceleration.

When the wave amplitude is below the critical value, CWRA does not initiate, and the modulational does not occurs, resulting in the electrons remaining at low energies. Since the maximum electron energy has not yet fully saturated at \(\omega_{pe}t = 447\) (Fig.~\ref{fig3}(a)), further studies should be performed using a larger system size.

%Beyond this point, a phase transition occurs in the $\psi'$-$\mu'$ space, where additional trapping regions emerge around $\psi=0$ and $2\pi$ (not shown here), and their separatrix size decreases with further increases in $\gamma'_{e}$. 
 
\section{Summary and Discussion}

In this study, we proposed a novel particle acceleration mechanism driven by large-amplitude Alfv\'en waves in a strong magnetic field. The acceleration process consists of multiple stages, including counter-propagating wave-particle resonant acceleration (CWRA) via decay instability, gyroresonant surfing acceleration (GRSA), and single wave resonant acceleration (SWRA). Our analysis demonstrates that these stages work in a sequential manner, leading to highly efficient energy gain for electrons. Initially, counter-propagating Alfvén waves resonantly accelerate electrons perpendicular to the ambient magnetic field via CWRA, which drives modulational instability and excites the modulational electrostatic field $E_{\text{mod}}$ within the wave packet. The ponderomotive force of the Alfvén wave packet then locally amplifies $E_{\text{mod}}$. Once $E_{\text{mod}}$ exceeds a certain threshold, it further enhances relativistic cyclotron resonance through GRSA. As the acceleration progresses, the $\mathbf{v} \times \mathbf{B}$ force facilitates the conversion of perpendicular momentum into parallel momentum. Eventually, when the electrostatic field diminishes, GRSA terminates, and electrons undergo continuous acceleration through SWRA, remaining trapped by the parent wave.

Additionally, the large $E_{\text{mod}}$ generates electrostatic trailing fields behind it, leading to trailing-field acceleration (TFA) of electrons. Our results indicate that the multi-stage acceleration process CWRA, followed by GRSA and SWRA, yields electron energies significantly exceeding those obtained through TFA alone. Notably, efficient acceleration occurs when the Alfv\'en wave amplitude exceeds the critical threshold for CWRA. By employing the test particle simulation in Section~III, the multi-stage acceleration process (CWRA, GRSA, and SWRA) is examined in detail through the analysis of electron trajectories in a monochromatic plane wave, in comparison with the results of the PIC simulation. It is well established that counter-propagating waves and modulational electrostatic fields naturally facilitate the injection of electrons into the high-energy wave trapping region.

The findings of this study have potential implications for various astrophysical environments where large-amplitude Alfv\'en waves exist. To date, various mechanisms have been proposed for the generation of high-energy cosmic rays, including magnetic reconnection \cite{Pino_1}, relativistic shocks \cite{Iwamoto_1}, and convective electric fields within relativistic plasma jets \cite{Mbarek_1}. The present study introduces the key concept of multi-stage resonant acceleration, along with TFA driven by Alfv\'en waves, potentially linked to these processes. Further refinement of this theory, including radiation effects, is required to more accurately estimate the maximum particle energy. In addition, future work should aim to explore the role of wave coherence and wave polarization, including multidimensional effects, in modifying the acceleration efficiency and the types of accelerated particles. Furthermore, the operation of these processes under realistic astrophysical conditions will be essential for refining acceleration models and interpreting observational data.

This work was performed under the joint research project of the Institute of Laser Engineering (ILE), the University of Osaka. S.I. was supported by the Japan Society for the Promotion of Science (JSPS) KAKENHI (Grants No. JP25K00221 and No. JP22K14020). T.S. was supported by the Japan Society for the Promotion of Science (JSPS) KAKENHI (Grants No. JP23K20038, No. JP21K03500, and No. JP20H00140), and by the JSPS Core-to-Core Program, B. Asia-Africa Science Platforms (Grant No. JPJSCCB20190003). S.M. was supported by the Japan Society for the Promotion of Science (JSPS) (Grant No. 23K22558) and by The Kajima Foundation for the International Joint Research Grants (2025-06). S.C. was supported by the National Science and Technology Council (NSTC), Taiwan (Grant No. 113-2112-M-008-010).

This is the accepted manuscript. The final published version is available at doi: https://doi.org/10.1103/nccj-mw1y

\section*{DATA AVAILABILITY}

The data that support the findings of this article are available from the corresponding author upon reasonable request.

\end{document}